\documentclass[twocolumn]{aastex61}
\pdfoutput=1 
\usepackage{amsmath,amstext}
\usepackage[T1]{fontenc}
\usepackage{apjfonts} 
\usepackage[figure,figure*]{hypcap}

\shorttitle{A Neutron Star Binary Merger Model for GW170817}
\shortauthors{Murguia-Berthier et al.}

\begin{document}
\title{A Neutron Star Binary Merger Model for GW170817/GRB170817a/SSS17a}

\author{A.~Murguia-Berthier}
\affiliation{Department of Astronomy and Astrophysics, University of California, Santa Cruz, CA 95064, USA}
\affiliation{DARK, Niels Bohr Institute, University of Copenhagen, Blegdamsvej 17, 2100 Copenhagen, Denmark}
\author{E.~Ramirez-Ruiz}
\affiliation{Department of Astronomy and Astrophysics, University of California, Santa Cruz, CA 95064, USA}
\affiliation{DARK, Niels Bohr Institute, University of Copenhagen, Blegdamsvej 17, 2100 Copenhagen, Denmark}
\author{C.~D.~Kilpatrick}
\affiliation{Department of Astronomy and Astrophysics, University of California, Santa Cruz, CA 95064, USA}
\author{R.~J.~Foley}
\affiliation{Department of Astronomy and Astrophysics, University of California, Santa Cruz, CA 95064, USA}
\author{D.~Kasen}
\affiliation{Nuclear Science Division, Lawrence Berkeley National Laboratory, Berkeley, CA 94720, USA}
\affiliation{Departments of Physics and Astronomy, University of California, Berkeley, CA 94720, USA}
\author{W.~H.~Lee}
\affiliation{Instituto de Astronom\'{i}a, Universidad Nacional Aut\'{o}noma de M\'{e}xico, Circuito Exterior, C.U., A. Postal 70-264, 04510 Cd. de M\'{e}xico, M\'{e}xico.}
\author{A.~L.~Piro}
\affiliation{The Observatories of the Carnegie Institution for Science, 813 Santa Barbara Street, Pasadena, CA 91101}
\author{D.~A.~Coulter}
\affiliation{Department of Astronomy and Astrophysics, University of California, Santa Cruz, CA 95064, USA}
\author{M.~R.~Drout}
\affiliation{The Observatories of the Carnegie Institution for Science, 813 Santa Barbara Street, Pasadena, CA 91101}
\author{B.~F.~Madore}
\affiliation{The Observatories of the Carnegie Institution for Science, 813 Santa Barbara Street, Pasadena, CA 91101}
\author{B.~J.~Shappee}
\affiliation{The Observatories of the Carnegie Institution for Science, 813 Santa Barbara Street, Pasadena, CA 91101}
\author{Y.-C.~Pan}
\affiliation{Department of Astronomy and Astrophysics, University of California, Santa Cruz, CA 95064, USA}
\author{J.~X.~Prochaska}
\affiliation{Department of Astronomy and Astrophysics, University of California, Santa Cruz, CA 95064, USA}
\author{A.~Rest}
\affiliation{Space Telescope Science Institute, 3700 San Martin Drive, Baltimore, MD 21218}
\affiliation{Department of Physics and Astronomy, The Johns Hopkins University, 3400 North Charles Street, Baltimore, MD 21218, USA} 
\author{C.~Rojas-Bravo}
\affiliation{Department of Astronomy and Astrophysics, University of California, Santa Cruz, CA 95064, USA}
\author{M.~R.~Siebert}
\affiliation{Department of Astronomy and Astrophysics, University of California, Santa Cruz, CA 95064, USA}
\author{J.~D.~Simon}
\affiliation{The Observatories of the Carnegie Institution for Science, 813 Santa Barbara Street, Pasadena, CA 91101}

\begin{abstract}
The merging neutron star gravitational wave event GW170817 has been observed throughout the entire electromagnetic spectrum from radio waves to $\gamma$-rays. The resulting energetics, variability, and light curves are shown to be consistent with GW170817 originating from the merger of two neutron stars, in all likelihood followed by the prompt gravitational collapse of the massive remnant. The available $\gamma$-ray, X-ray and radio data provide a clear probe for the nature of the relativistic ejecta and the non-thermal processes occurring within, while the ultraviolet, optical and infrared emission are shown to probe material torn during the merger and subsequently heated by the decay of freshly synthesized $r$-process material. The simplest hypothesis that the non-thermal emission is due to a low-luminosity short $\gamma$-ray burst (sGRB) seems to agree with the present data. While low luminosity sGRBs might be common, we show here that the collective prompt and multi-wavelength observations are also consistent with a typical, powerful sGRB seen off-axis. Detailed follow-up observations are thus essential before we can place stringent constraints on the nature of the relativistic ejecta in GW170817.
\end{abstract}

\section{Introduction}
The discovery of galactic binary neutron stars \citep{Hulse1975} firmly established the existence of a class of systems which would merge in less than a Hubble time via the emission of gravitational wave emission. Over the years, various studies showed that these binaries are in principle capable of powering cosmological $\gamma$-ray bursts of the short variety  \citep{Kouveliotou1993} when they merge \citep{Paczynski1986,1992ApJ...395L..83N, Eichler1989,Piran_review,Lee07,Gehrels09,2015PhR...561....1K}, while those of the long variety have been shown to be associated to the core collapse of massive stars \citep{WoosleyBloom2006}. After decades of instrumental, observational and theoretical progress, a watershed event occurred on 17 August 2017, when the Swope Supernova Survey discovered the first optical counterpart of a gravitational wave event, GW170817, attributed to the merger of two neutron stars \citep{GCN21509}, named SSS17a \citep{Coulter17}. This detection led to the measurement of a redshift distance and thus the firm identification of the candidate host galaxy, NGC~4993 \citep{Coulter17, Pan17}. 

One of the key electromagnetic discoveries concerned the detection of GRB~170817a by {\it Fermi} and INTEGRAL \citep{fermi,integral}, a short $\gamma$-ray burst (sGRB) lasting only a few tenths of seconds. The short duration and dim signal of the prompt $\gamma$-ray transient, however, precluded the determination of an accurate position until the Swope Supernova Survey succeeded in promptly localizing GRB~170817a/SSS17a \citep{fermi, integral, Coulter17}. For the next few days, several multi wavelength
observations were made \citep{integral, chandra, vla, capstone}. The detection of GW170817 and follow-up electromagnetic observations have revolutionized 
our view of merging neutron stars, confirming some previously held ideas and adding invaluable elements to our knowledge of them. The concept of a sudden release of energy almost exclusively concentrated in a brief pulse of $\gamma$-rays has
been discarded. Indeed, even the term {\it afterglow} should be now recognized as misleading as the energy radiated during the first three weeks at longer wavelengths greatly exceeds that emitted during the prompt $\gamma$-ray phase \citep{fermi}. 

The broad electromagnetic manifestations of GW170817 thus provide us with a unique opportunity, to which this {\it Letter} is dedicated, to constrain the ejecta properties following the merger of a binary neutron star. In Section \ref{sec:meta} we address
the energetics and timescales of the observed radiation and compare it with the data from sGRBs. In Sections \ref{sec:lowl} and \ref{sec:offaxis} we constrain the properties of the ejecta by using all the information available to us from both the afterglow and prompt radiation. We summarize 
our findings in Section \ref{sec:dis}.

\section{Metabolics of GW170817/SSS17a}\label{sec:meta}
Here we construct a basic inventory of the energy radiated at all wave bands from $\gamma$-rays to radio waves using all data collected for the GW170817 event by the One-Meter Two-Hemisphere (1M2H) collaboration \citep{Coulter17, Drout17, Kilpatrick17, Shappee17, Siebert17} as well as from other publicly available sources \citep{vla,chandra,integral}.
Such a basic inventory provides an assessment of the various modes of energy transfer and release involved within the ejecta. The compilation also offers a way to test our
understanding of the physics of neutron star mergers. Some apparent points should be emphasized. We
measure directly only the the energy radiated in the direction of 
the Earth per second per steradian per frequency interval by the source. The apparent bolometric luminosity $L_{\rm iso}$ may be quite different from the true bolometric luminosity if the
source is not isotropic. To investigate the energy dissipation history we fitted a natural cubic spline function to the luminosity $L_{\rm iso}$
as a function of time at different frequency intervals. This allows us to estimate the cumulative
emitted energy $E_{\rm iso}$ and also derive $t_{\rm 90}$ values at all energies, which we define here as the time in the source frame
during which 90\% of the radiated energy is accumulated. Using this, we derive $E_{\rm iso}$ and $t_{\rm 90}$ at $\gamma$-ray (75-2000 keV) from \citealp{capstone}, X-ray (0.3-10 keV), ultraviolet (2600-3465 \AA), optical (3465-9665 \AA), infrared (10200-21900 \AA) and radio (5-10 GHz) energies:
\begin{itemize}
\item $E_{\rm \gamma, iso} \approx 5.3\pm1.1 \times 10^{46}$ erg and $t_{\rm 90, \gamma}\approx 2$ s
\item $E_{\rm IR,iso } \approx 2.7\pm 0.5 \times 10^{48}$ erg and $t_{\rm 90, IR}\approx 10.5$ days
\item $E_{\rm O, iso} \approx 4.1\pm 0.4 \times 10^{47}$ erg and $t_{\rm 90, O}\approx 3.8$ days
\item $E_{\rm UV, iso} \approx 1.3\pm 0.6 \times 10^{47}$ erg and $t_{\rm 90, UV}\approx 1.1$ days
\item $E_{\rm X, iso} \approx 1.9\pm 0.5 \times 10^{44}$ erg and $t_{\rm 90, X} \gtrsim 15.2$ days
\item $E_{\rm R, iso} \lesssim 8 \times 10^{40}$ erg and $t_{\rm 90, R} \gtrsim 17.7$ days
\end{itemize}

 \begin{figure}
\centering
\includegraphics[width=0.5\textwidth]{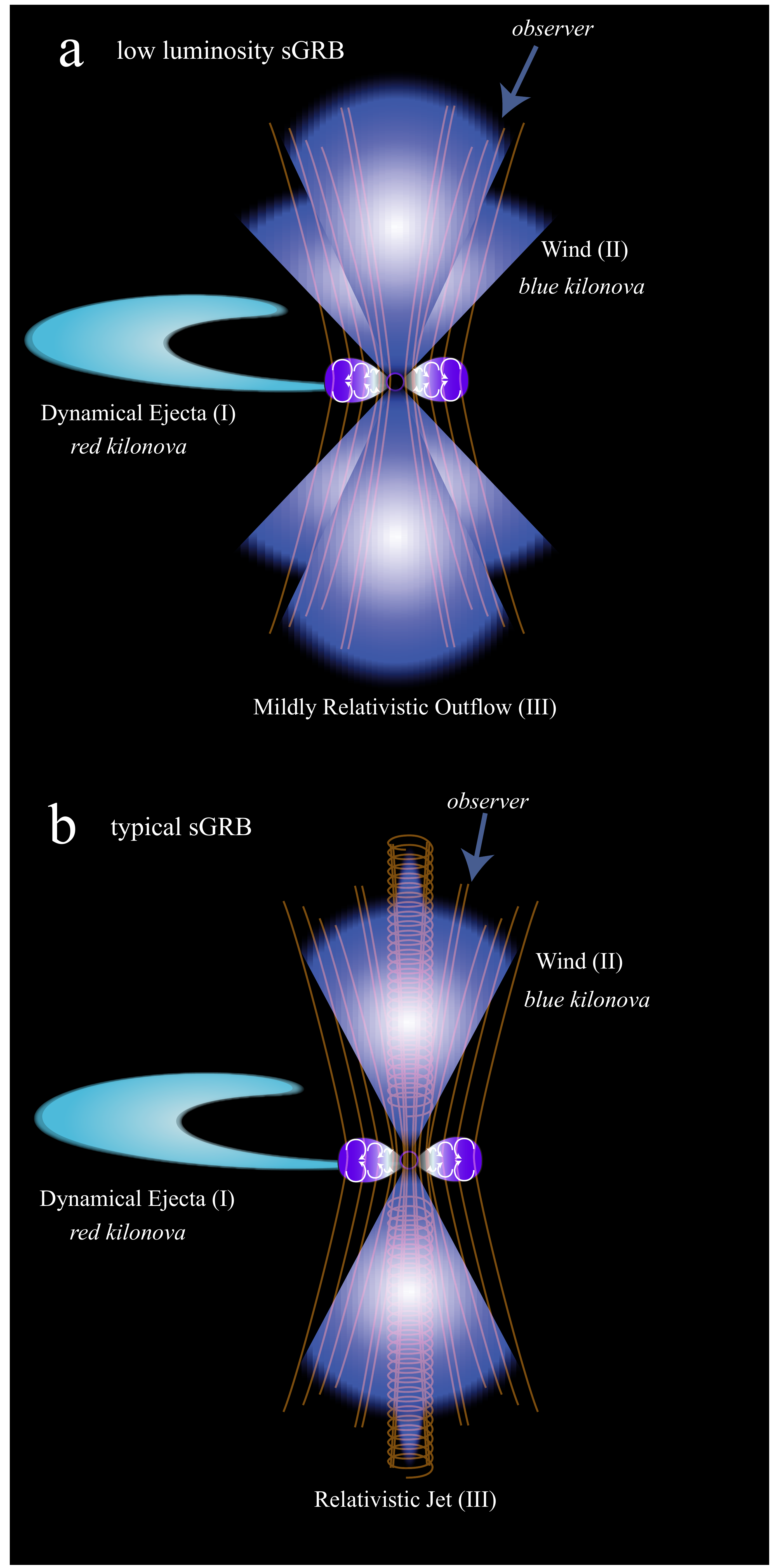}
\caption{An overview of the main energy transfer processes thought to be involved in ejecting material in neutron star mergers. As
they merge a few percent of the matter is ejected in the form of a tidal tail (I). The shocked merged remnant is expected to produce strong winds (II) and is likely to be top-heavy and unable to survive. The expected outcome is the collapse to a black hole. A spinning black hole constitutes an excellent gyroscope, and the ingredients of accretion and magnetic fields are probably sufficient to
ensure the production of a sGRB jet (III in scenario {\it b}). A potential death-trap for such highly relativistic outflows is the amount of entrained baryonic mass, which can severely limit their power (III in scenario {\it a}). }
\label{fig1}
\end{figure}

While unremarkable for its duration, GRB~170817a had a total energy 
 that is some 4-6 orders of magnitude less than a
typical {\it Swift} sGRB \citep{Gehrels09}.
The currently inferred isotropic X-ray emission $E_{\rm iso, X}$, which for most {\it Swift} sGRBs is comparable to that emitted during the prompt $\gamma$-ray phase, is at least 6-8 orders 
of magnitude smaller \citep{Gehrels09}. 
The isotropic equivalent energy that is radiated at optical wavelengths
in this case is 2 orders of magnitude larger than that in $\gamma$-rays. This is in stark contrast to {\it Swift} sGRBs, for which $E_{\rm O, iso}$ is at least 2 orders of magnitude smaller than $E_{\rm \gamma, iso}$. What is more, GW170817/SSS17a is radically different in its optical properties from any other known sGRBs \citep{Siebert17}. 
 The optical emission rises in less than half a day, then fades rapidly, exhibiting a swift color evolution to redder wavelengths \citep{Drout17}. 
 While optical sGRB afterglows can produce rapidly fading transients, they don't generate the quasi-blackbody spectrum that is observed in SSS17a \citep{Shappee17}.
These results are consistent with the emerging hypothesis that the ultraviolet, optical and infrared 
emission probe matter torn from the merger system, ejected at sub-relativistic velocities and subsequently heated by the decay of freshly synthesized $r$-process material \citep{Kilpatrick17,Kasen17}.
 
On its own, the low-luminosity $\gamma$-ray emission of the unusually faint GRB~170817a can thus support the idea
of a common class of intrinsically sub-energetic sGRBs. The key question is whether 
there is significant observational
support for the existence of low amounts of relativistic energy released during this event or whether the afterglow light curves are instead more consistent with a
model in which GRB~170817a was a classical jetted
sGRB viewed off-axis. Figure \ref{fig1} presents here our selections for the
energy transfer channels during merging neutron star binaries that we believe are responsible for the various entries in the inventory. Material dynamically stripped during the merger (denoted I in Figure \ref{fig1}) is ejected by tidal torques through the outer Lagrange point, removing energy and angular
momentum and forming a large tidal tail \citep{1998ApJ...494L..53K,Rosswog05, 2012LRR....15....8F}. 
This material is expected to undergo $r$-process
nucleosynthesis and give rise to a red (quasi-thermal) kilonova \citep{Li98,1999ApJ...525L.121F, Metzger10, Roberts11}. 
The configuration after merger consists
of a hyper massive neutron star (HMNS, one with more mass than a cold, non-rotating
configuration could support) surrounded by an extended shock-heated envelope \citep{2000ApJ...528L..29B,2006PhRvL..96c1101D}. During this stage, various dissipation and
transport mechanisms can give rise  \citep{Perego14, Siegel14} to strong winds (denoted II in Figure \ref{fig1}). These are thought to produce low-opacity (first-peak) $r$-process material, giving rise to a blue (quasi-thermal) kilonova \citep{Kasen13,Kasen15,Metzger16}. 
The properties of the HMNS have
a decisive outcome on whether or not a standard sGRB will be observed \citep{ari14,ari17, 2017ApJ...844L..19P}. This is because even a tiny
mass of baryons polluting the jet will severely
limit the maximum attainable Lorentz factor and 
effective jet triggering might have to wait until after black hole collapse.
In scenario {\it a}, the wind emanating from the HMNS hampers the advancement of a relativistic jet, leading to a low luminosity event \citep{2002MNRAS.336L...7R, 2003MNRAS.343L..36R, 2014ApJ...784L..28N, 2016ApJ...816L..30J}. 
In scenario {\it b} in Figure \ref{fig1}, the collapse to a black hole occurs promptly and a classical jetted sGRB is produced which we happen to view off-axis (such as \citealp{Enrico05}).
Not only would a sGRB be detectable in both scenarios, followed by an afterglow, but there could also be additional extended emission at early stages
caused by the reprocessing of this energy and its subsequent dissipation \citep{ari14,2015ApJ...802...95R}.
This could
resemble the so called extended emission in sGRBs \citep{2006ApJ...643..266N}.
In the following sections, we examine
these two possible interpretations and critically asses whether the electromagnetic observations of 
GW170817/SSS17a support the idea that it was an intrinsically weak, nearly isotropic explosion or either a classical sGRB, such as GRB 130603b \citep{fong15},
observed
off-axis. An accurate assessment of the kinetic energy content in relativistic material requires detailed afterglow modeling.

\section{A Low Luminosity sGRB}\label{sec:lowl}
GW170817/SSS17a/GRB~170817a, or at least the $\gamma$-ray emission along our line of sight, was
certainly feeble. The simplest interpretation might be
that the $\gamma$-ray emission was deficient in all directions (scenario {\it a} in Figure~\ref{fig1}), as in the case of low luminosity long GRBs associated with type Ic supernovae \citep{Kaneko07}.

 \begin{figure}
\centering
\includegraphics[width=0.485\textwidth]{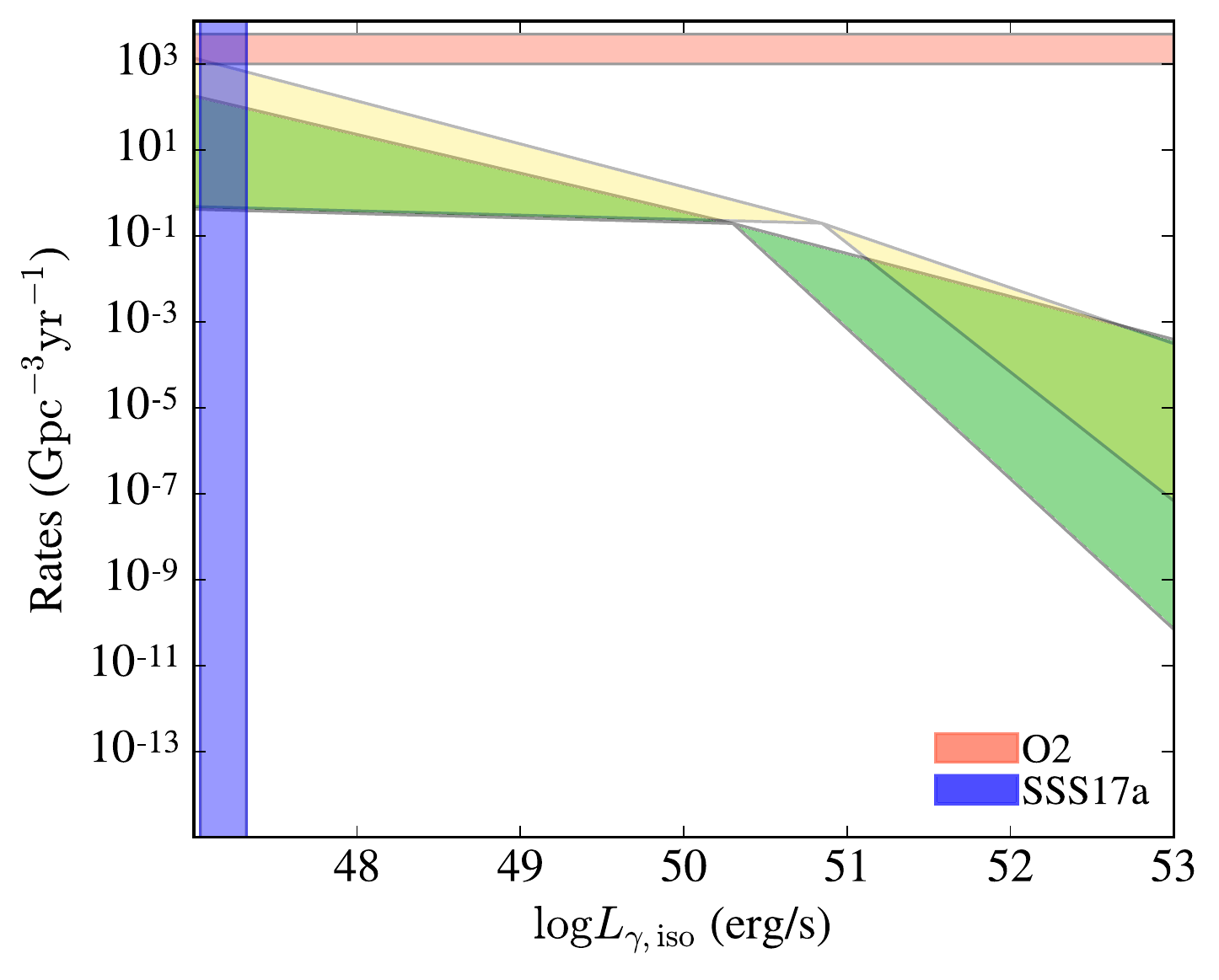}
\caption{Shown are two luminosity functions taken from \citealp{Guetta06}. They are described by a broken power-law peak luminosity function with $L_\ast=0.2\times10^{51}\ {\rm erg/s}$, $a=0.6^{+0.3}_{-0.5}$, $b=1.5^{+2}_{-0.5}$ (green) and $L_\ast=0.7\times10^{51}\ {\rm erg/s}$, $a=0.6^{+0.4}_{-0.5}$, $b=2^{+1}_{-0.7}$ (yellow). If we assume
a beaming correction factor of $27^{+158}_{-18}$ we find a merger rate that is broadly consistent with estimated O2 LIGO rates \citealp{Ligorate} and can accommodate the $L_{\rm \gamma, iso}$ measured by \citealp{fermi} for GW170817/SSS17a/GRB~170817a.}
\label{fig2}
\end{figure}

Such a weak burst, thousands to millions of times fainter than the inferred isotropic energies of sGRB, could belong to a separate population of weakly jetted, low luminosity events. We thus need to quantify the odds of detecting such an event as non-Euclidean number count statistics limit the fraction of bursts that can be observed from the local Universe \citep{1998ApJ...506L.105B}.

In Figure~\ref{fig2} we compare the properties of GRB~170817a \citep{fermi} with the luminosity function of sGRBs as constrained from the peak flux distribution of BATSE events and the 
redshift and luminosity distributions of {\it Swift} events. The luminosity functions shown in Figure~\ref{fig2} have been derived \citep{Guetta06} under the assumption that the sGRB rate follows a distribution of delay times that is consistent with those commonly used to describe the merging rate of double neutron star binaries \citep{2004MNRAS.350L..61C, 2014ApJ...792..123B, 2015ApJ...807..115S}. 

If we assume a typical beaming correction of $27^{+158}_{-18}$ for sGRBs \citep{fong15}, we find an event rate that is broadly consistent with the O2 LIGO merger rate estimated by \citealp{Ligorate} and can accommodate the $L_{\rm \gamma, iso}$ measured by \citealp{fermi}, under the assumption that GRB~170817a was similarly weak in all directions. BATSE was a benchmark experiment that produced a catalogue containing more than 2,000 GRBs \citep{1999ApJS..122..465P}. How many of these bursts could have been GW170817-like events?
The observed number of sGRBs and the lack of excess events from the direction of the Virgo cluster suggests that only a tiny fraction ($\lesssim 0.05$) of these bursts can be like GW170817 within $\lesssim 40$ Mpc \citep{2005Natur.434.1107P}.

\subsection{Prompt Emission}\label{sec:prompt}
The energy spectrum for GRB170817a is well described by a power law with an exponential cutoff at $\approx 185\ {\rm keV}$ \citep{fermi}. With no significant emission observed above 300 keV, GRB170817a is an example of the {\it no high-energy} bursts that compose 25\% of the BATSE sample \citep{1999ApJS..122..465P}. Since GRB~170817a had a single-peaked 
light curve \citep{fermi}, the burst variability, $\delta t_{\rm var}$, is roughly given by $ \delta t_{\rm var} \approx t_{\rm 90, \gamma} \approx 2 \pm 0.5$s \citep{capstone}.

A constraint on the size of the emitting region $R_\gamma$ can be derived from the delay time $\delta t_{\rm gw} \approx  t_{\rm 90, \gamma} \approx 2$s \citep{capstone} observed between the arrival of the prompt $\gamma$-ray emission and the gravitational wave merger signal. If one assumes that the relativistic outflow, moving at $\Gamma=(1-\beta^2)^{-1/2}$, was triggered at merger, then 
\begin{equation*}
R_\gamma =c \delta t_{\rm gw} \beta (\beta-1)^{-1} \approx 2 \Gamma^2 c \delta t_{\rm gw}, 
\end{equation*}
which is consistent with most studies aimed at understanding the nature of the $\gamma$-ray dissipation in sGRBs \citep{2007PhR...442..166N,2015PhR...561....1K}. Internal shocks dissipation, for example, is thought to occur at a radius $R_{\iota} \approx 2 \Gamma^2 c \delta t_{\rm v}$ \citep{1992MNRAS.258P..41R}. Since $\delta t_{\rm var} \approx \delta t_{\rm gw}$, it follows that $R_{\iota}\approx R_\gamma$.

\subsection{The Afterglow Emission}
If there was indeed this amount of relativistic energy we
can then try to explain why we did not see the afterglow emission at early times
by invoking a standard afterglow model. In such quasi-spherical case, the
emission we expect should be below the ultraviolet, optical and infrared emission, which is dominated by heating from the decay of freshly synthesized $r$-process material \citep{Coulter17, Drout17,Kilpatrick17,Shappee17, Siebert17}.

 \begin{figure}
\centering
\includegraphics[width=0.43\textwidth]{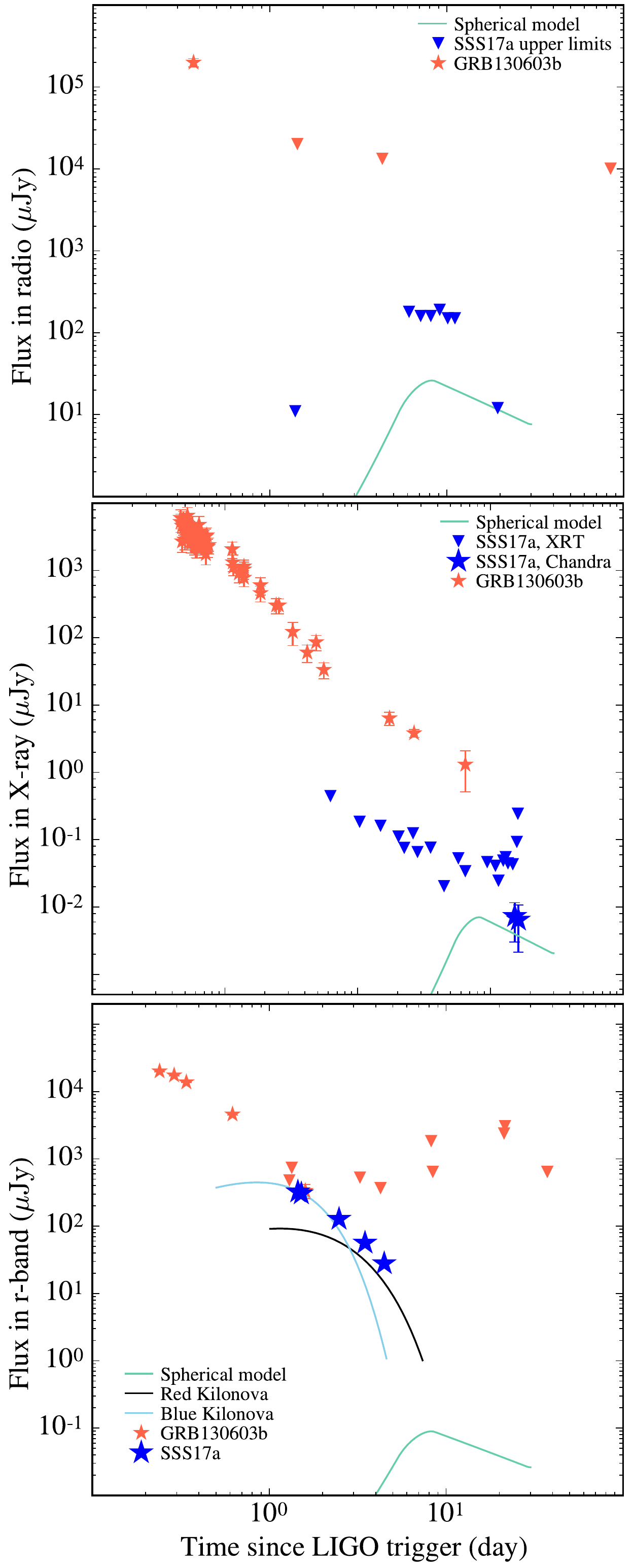}
\caption{Afterglow emission from a spherical, low energy blast wave (scenario {\it a} in Figure~\ref{fig1}) at optical (r-band), radio (6 GHz), and X-rays ($7 \times 10^{16}$ Hz). 
The afterglow light curves presented here are calculated using the blast wave models from \citealp{spherefit}. The microphysical parameters, the energetics and the properties of the external medium and burst energetics are given in the text. Also plotted is data from the One-Meter Two-Hemisphere collaboration: \citealp{Coulter17, Drout17,Kilpatrick17,Shappee17, Siebert17} for the r-band, \citealp{chandra} for the X-rays, and \citealp{vla}, EuroVLBI team for the radio. The afterglow detections and upper-limits of the standard GRB~130603b are plotted for comparison \citep{fong15}.}
\label{fig3}
\end{figure}

The resulting light curves for a low energy spherical model
are plotted against observations of SSS17a in Figure~\ref{fig3}. As a point of comparison, we plot the light curves of GRB~130603b, whose afterglow properties are representative of classical sGRBs \citep{fong15}.
The local emissivity is calculated using standard assumptions of synchrotron emission from
relativistic electrons that are accelerated behind the shock (with a power-law distribution of energies wind index $p$)
where the magnetic field and the accelerated electrons hold fractions $\epsilon_e$ and $\epsilon_B$, respectively, of the internal energy.The model parameters are: $n=0.08\ {\rm cm^{-3}}$; $E_{\rm K, iso}=8\times 10^{48}\ {\rm erg}$, $p=2.1$, $\epsilon_B=0.05$, $\epsilon_e=0.05$ and the fraction of electrons that get accelerated is $\xi_N=1$. We emphasize that the model parameters cannot be uniquely determined from the fit to the multi-wavelength  observations, and other sets of parameters could provide an equally acceptable description. The X-ray emission is very sparsely sampled and thus provide only  mild constraints on models on its own. However, when combined with the optical and radio limits, they provide a better handle on the model, thus significantly improving upon the constraints derived from the X-ray data alone. The reader is refer to \citealp{2015PhR...561....1K} for a detailed description of our current understanding regarding afterglow physics and observational constraints.

The fact that X-ray emission was seen at $t=15.2$ days, implies that $t_{\rm dec}\lesssim$ 15.2 days, where $t_{\rm dec}=R_{\rm dec}/(2c\Gamma^2)$ and $R_{\rm dec}=(3E_{\rm k,iso}/4\pi n m_pc^2\Gamma^2)^{1/3}$ are the observed time and radius at which the outflow  decelerates  appreciably \citep{Piran_review}. In the model depicted in Figure~\ref{fig3}, the initial Lorentz factor of the blast wave is chosen to be $\Gamma = 5.5$ in order for $t_{\rm dec}\lesssim$ 15.2 days. In this case, $R_\gamma \approx 3.6 \times 10^{12} (\Gamma/5.5)^2 (\delta t_{\rm gw}/ 2{\rm s})$ cm.

Some points from should be emphasized here. The afterglow light curves provide a reasonable description of the sparsely sampled X-ray afterglow and are consistent with the lack of non-thermal radiation observed at radio and optical wavelengths. The optical emission is dominated by quasi-thermal emission, which also dominates the total radiated energy output \citep{Drout17}. This is illustrated in Figure~\ref{fig3}, where the best fit models for the kilonova emission at optical wavelengths are plotted. These models have been constructed using the simple formalism developed by \citealp{2017LRR....20....3M} and are tailored to match the values derived in \citealp{Kilpatrick17} using more sophisticated models. In this simple model, we contemplate ejecta of mass $m_{\rm ejecta}$ expanding at a velocity $v_{\rm ejecta}$, which is heated by the decay of freshly synthesized $r$-process material. Two different ingredients are assumed for the ejecta: a {\it blue} ($m_{\rm ejecta}=0.025M_\odot$ and $v_{\rm ejecta}=0.3c$) and a {\it red} ($m_{\rm ejecta}=0.035M_\odot$ and $v_{\rm ejecta}=0.15c$) component. We use $\kappa_{\rm blue}=0.08\ \rm{cm^2g^{-1}}$ and $\kappa_{\rm red}=5\ \rm{cm^2g^{-1}}$ to describe the opacity of the blue (lanthanide free) and red (lanthanide rich) components, respectively \citep{2013ApJ...775...18B}. This two component model, as argued in \citealp{Kilpatrick17}, is in remarkable agreement with the wealth of observations our team has assembled at optical, ultraviolet and infrared wavelengths \citep{Coulter17, Drout17,Kilpatrick17,Shappee17, Siebert17}.
What is more, observations at $\gamma$-rays, X-rays and radio wavelengths are consistent with GRB~170817a being an intrinsically weak, nearly isotropic explosion. Having said this, continuous monitoring of the source at X-ray and radio wavelengths could render this type of model
unacceptable if the integrated energy is observed to increase.

\section{An off-axis Model}\label{sec:offaxis}
Given that most sGRBs are collimated \citep{fong15},
 their observed properties will unavoidably change depending
on the angle $\theta_{\rm obs}$ (measured with respect to the jet axis) at which
they are observed. If we make the standard assumption of a top-hat jet,
the prompt and afterglow properties of the sGRB would be almost the same to all observers located within the initial jet aperture, 
denoted here as $\theta_0$. At $\theta_{\rm obs} > \theta_0$, the jet emission is expected to decline precipitously \citep{Granot02b}. 
 
 \subsection{Prompt Emission}
\begin{figure}
\centering 
\includegraphics[width=0.485\textwidth]{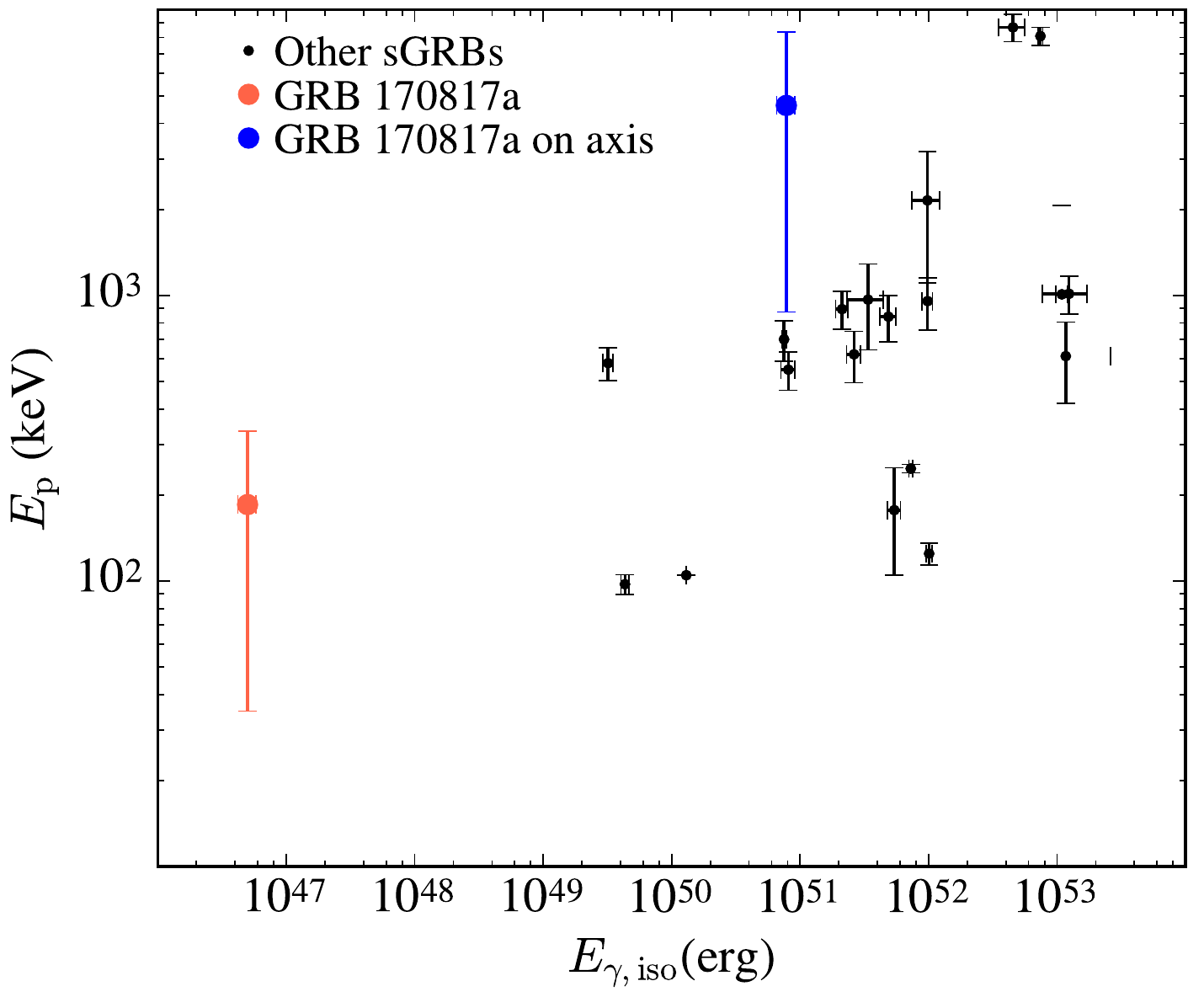}
\caption{ The location of GRB~170817a in the $E_{\rm p}$ and $E_{\rm \gamma,iso}$ plane, from \citealp{integral, fermi}. Also shown is the location if GRB~170817a were on-axis under the assumption of a misaligned, sharp-edged jet. This assumes a Lorentz factor of $\Gamma\approx 50$ and $\Gamma(\theta_{\rm obs}-\theta_0) \approx 5$ (Section~\ref{sec:ofaft}). The data for the other sGRBs are taken from \citealp{Tsutsui13} and \citealp{Avanzo14}.}  
\label{fig4}
\end{figure}

 In a typical sGRB (scenario {\it b} in Figure~\ref{fig1}), the $\gamma$-rays we detect are concentrated into a cone of opening angle comparable to $\theta_0$, provided that $\theta_0> \Gamma^{-1}$. Thus, if the jet is viewed at $\theta_{\rm obs}>\theta_0$ from the jet axis, the $\gamma$-ray luminosity will be drastically suppressed. For a jet with $\Gamma$, the typical peak photon
energy $E_{\rm p}$ scales as \citep{Enrico05}
 \begin{equation*}
 E_{\rm p}\propto [\Gamma(\theta_{\rm obs}-\theta_0)]^{-2}, 
 \end{equation*}
 while 
 \begin{equation*}
 E_{\rm \gamma, iso}\propto [\Gamma(\theta_{\rm obs}-\theta_0)]^{-6}.
 \end{equation*}
Figure~\ref{fig4} shows a sample of observed $E_{\rm p}$ and $E_{\rm \gamma,iso}$ for sGRBs together with the properties of GRB~170817a if it were viewed on-axis.  In order to generate the on-axis conditions we have used $\theta_{\rm obs}=1.5\theta_0$ and $\theta_0=0.2$, which we inferred from a fit to the afterglow emission (see Section~\ref{sec:ofaft} for details), and have assumed $\Gamma=50$. These values are compatible with those observed in (and in some cases derived for) sGRBs \citep{Gehrels09,Berger14, fong15}. We thus consider $\Gamma=50$ and the inferred
on-axis values of $E_{\rm p} \approx 4$ MeV and $E_{\rm \gamma,iso} \approx 7\times 10^{50}\ {\rm erg}$ to be reasonable for conditions expected at the edge of the jet \citep{Enrico05}.

For $\theta_{\rm obs}>\theta_0$, one expects some significant decrease in the variability of the prompt emission. This is because the duration of an individual pulse in the light curve scales as $\delta t_{\rm var} \propto [\Gamma(\theta_{\rm obs}-\theta_0)]^{2}$. Since the distance between neighboring pulses is typically comparable
to the width of an individual pulse,  then a sizable increase in $\delta t_{\rm v}$ could cause
significant overlap between pulses and, as a result, the variability would be washed out. 

The total duration of the event could also  increase significantly for large viewing angles when $[\Gamma(\theta_{\rm obs}-\theta_0)]^{2} \gtrsim (t_{\rm 90, \gamma} / \delta t_{\rm var})$, where the total observed duration of the burst ($t_{\rm 90, \gamma}$) and individual pulse variability ($\delta t_{\rm var}$) are measured  here on-axis. For most sGRBs, ($t_{\rm 90, \gamma} / \delta t_{\rm var})\approx 10-10^2$ \citep{Gehrels09}, which implies that for $[\Gamma(\theta_{\rm obs}-\theta_0)]^{2} < t_{\rm 90, \gamma} / \delta t_{\rm var} \sim 10-10^2$ the total duration of the burst when observed off-axis should not increase significantly. The variability of the burst, when observed off-axis on the other hand, is expected to be smeared out. 

Based on the model parameters estimated here, we thus expect GRB170817a to have been significantly more luminous, have a shorter duration, be more variable and have a much harder spectrum for observers located within $\theta_{\rm obs} \lesssim \theta_0$. This might explain why GRB170817a was observed to be somewhat less variable than typical sGRBs. As argued in Section~\ref{sec:prompt}, the observed time delay between the arrival of the gravitational wave signal and the prompt $\gamma$-ray emission can be used to place constraints on the size of the emitting region although the degree of pulse and light curve smearing  in this scenario complicates the calculation.

 \begin{figure}
\centering
\includegraphics[width=0.43\textwidth]{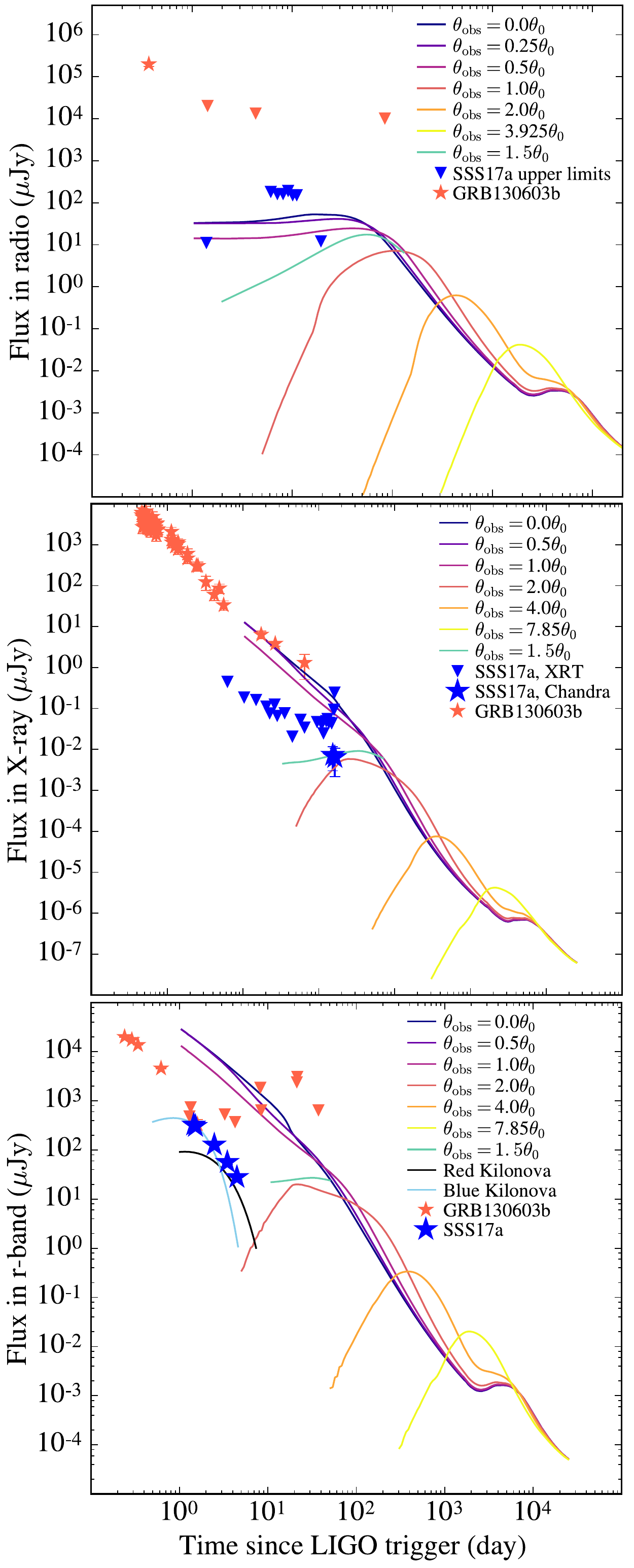}
\caption{ Afterglow emission from a standard off-axis jet (scenario {\it b} in Figure~\ref{fig1}). Light curves are calculated for various viewing angles $\theta_{\rm obs}$ at optical (r-band), radio (6 GHz), and X-rays ($7 \times 10^{16}$ Hz) for a sGRB with standard parameters: $n=0.3\ {\rm cm^{-3}}$, $E_{\rm K, iso}=2.5\times 10^{51}\ {\rm erg}$, $p=2.8$, $\epsilon_B=0.002$, $\epsilon_e=0.02$ and $\xi_N=1$. The curves presented here are calculated using the models from \citet{vanEerten11}. Also plotted is the same data and kilonova models shown in Figure~\ref{fig3}. The data 
for GRB~170817a/SSS17a can be reasonably fit by a standard sGRB seen seen at $\theta_{\rm obs}=1.5\theta_0$, where $\theta_0$=0.2. The on axis model is broadly compatible with the properties of typical sGRBs as illustrated by the comparison with GRB~130603b \citep{fong15}.}
\label{fig5}
\end{figure}

\subsection{Afterglow Emission}\label{sec:ofaft}

An observer at $\theta_{\rm obs} > \theta_0$ observes a rising afterglow light curve at early times \citep{Granot02b}. The afterglow light curve will be observed to peak when $\Gamma$, which is decreasing with time, reaches a value $\approx (\theta_{\rm obs}-\theta_0)^{-1}$ and soon after will approach that seen by an on-axis observer. This can be discerned 
by comparing the curves for $\theta_{\rm obs}=\theta_0$ and $\theta_{\rm obs}=2\theta_0$ in Figure~\ref{fig5}. The observations of GRB~170817a/SSS17a can be accommodated if $\theta_{\rm obs}\approx 1.5 \theta_0$ for $\theta_0=0.2$. That is, our line of sight happened to be a few degrees from a sharp-edged typical sGRB jet. The constraints inflicted by the properties of the afterglow 
emission thus support the idea that GRB~170817a/SSS17a was a standard sGRB
jet seen off-axis. The on-axis model $\theta_{\rm obs}\lesssim \theta_0$ in this case provides a reasonable description of the broad afterglow properties of GRB~130603b \citep{fong15}, which by all accounts is representative of the sGRB population.
The isotropic kinetic energy of the jet, when viewed on axis, would be $E_{\rm K, iso}=2.5\times 10^{51}\ {\rm erg}$, which can explain (for reasonable dissipation efficiencies) $E_{\rm \gamma,iso} (\theta_{\rm obs}\lesssim \theta_0)$ in Figure~\ref{fig4}. The simple off-axis fit to the afterglow observations does not, however,
uniquely determine the model parameters. Some model constraints are, however, rather robust. Most notably within the framework presented here, 
if the jet axis had been closer to the observer's direction, the intensity of the optical and infrared afterglow might have
prevented us from uncovering the kilonova signal. This can be clearly seen by comparing the properties of SSS17a with those of GRB~130603b. This implies that the edge of the jet must be sufficiently
sharp, so that the emission at early times would be dominated by the core of the jet, rather than by material along the line of sight that might produce bright radio, optical and X-ray emission.

\section{CONCLUSION AND PROSPECTS}\label{sec:dis}
The recent discovery of GRB~170817a/SSS17a associated with GW170817 \citep{Coulter17,fermi} has made it possible to strengthen the
case for binary neutron star mergers as the main progenitors of sGRBs \citep{Piran_review, Lee07, 2007PhR...442..166N, Gehrels09, 2011ApJ...732L...6R, Berger14}.
While the isotropic energy emitted in gravitational waves is of the order of a
fraction of a solar rest mass $\gtrsim 0.025 M_\odot c^2\approx 4.5 \times 10^{52}$ erg \citep{gw_paper}, the total integrated electromagnetic emission is estimated here to be drastically lower $\approx 5 \times 10^{48}$ erg (Section~\ref{sec:meta}) and was dominated by 
the quasi-thermal emission seen at infrared, optical and ultraviolet wavelengths. By modeling this emission in great detail, \citealp{Kilpatrick17} predicts the kinetic ejecta content at sub-relativistic velocities ($v_{\rm ejecta}\approx0.1c$) to be of the order of a few times $10^{51}$ erg. 

The faint nature of GRB~170817a can be used to
argue for the existence of at least two different possibilities for the nature of the sGRB event associated with GW170817, on the basis of different amounts of relativistic energy released
during the initial explosion. In this {\it Letter}, we have examined 
two concrete alternatives.  The first one is based on the premise
 that GRB~170817a was an intrinsically weak, nearly isotropic explosion and we conclude that current observations are also consistent
with this idea (Section~\ref{sec:lowl}). In this scenario, the kinetic energy content at mildly relativistic velocities $\Gamma \approx 5$ is estimated to be $\approx10^{49}$ erg.
 The second alternative is based on the hypothesis that GRB~170817a was an ordinary GRB observed
off-axis, and in this case we conclude that current available data is consistent with an off-axis model in which GRB~170817a was a
much more powerful event seen at an angle of about 1.5 times
the opening angle of the jet (Section~\ref{sec:offaxis}). The kinetic energy content at relativistic velocities $\Gamma\approx10^2$ is thus estimated to be $\approx10^{50}$ erg after correcting for beaming.
 Detailed X-ray and radio follow-up observations and polarimetry of GRB~170817a/SSS17a should provide us with stringent
constraints on the jet geometry and energetics, as both models make very different predictions. \footnote{It should be noted that there is a disagreement about the possible detection of a radio afterglow at $t=$17-19 days \citep{GCN21815,vla}. If confirmed both models presented here can be slightly modified to provide a reasonable description of the data at this specific epoch.} The off-axis model, for example, will be preferred if the X-ray and radio fluxes are observed to increase.

The progenitors of sGRBs have been until now essentially masked by afterglow
emission, which is largely featureless synchrotron emission \citep{2007PhR...442..166N}. The detection of kilonova emission
has clearly established the potential of electromagnetic signatures to shed light on the properties of the ejecta and its composition after merger \citep{Kilpatrick17}. As we have described, our rationalization of the principal post-merger physical considerations combines some generally accepted principles with some more speculative 
ingredients (Figure~\ref{fig1}). When confronted with observations, it seems to accommodate the gross properties 
of the electromagnetic radiation (Figures~\ref{fig3} and \ref{fig5}), in addition to the incalculable value of the information that will be gathered from the concurrent gravitational
event will provide us with the exciting opportunity to study and test new regimes of physics. 

\section*{Acknowledgments}
We thank the LIGO/Virgo Collaboration, and all those who have contributed to gravitational wave science for enabling this discovery.
We thank J.\ McIver and B.\ Mockler, and the anonymous referee.
We would like to thank I.\ Thompson, J.\ Mulchaey (Carnegie), L.\ Infante and the entire Las Campanas staff.
We thank K.\ Alexander, W-F.\ Fong, R.\ Margutti , the EuroVLBI team, the INTEGRAL team, the Chandra team and the Fermi-GBM team for granting permission to use their data. 
We thank the University of Copenhagen, DARK Cosmology Centre, and the Niels Bohr International Academy for hosting D.A.C., R.J.F., A.M.B., E.R., and M.R.S.\ during the discovery of GW170817/SSS17a. R.J.F., A.M.B., and E.R.\ were participating in the Kavli Summer Program in Astrophysics, ``Astrophysics with gravitational wave detections.'' This program was supported by the the Kavli Foundation, Danish National Research Foundation, the Niels Bohr International Academy, and the DARK Cosmology Centre.
The UCSC group is supported in part by NSF grant AST--1518052, the Gordon \& Betty Moore Foundation, the Heising-Simons Foundation, generous donations from many individuals through a UCSC Giving Day grant, and from fellowships from the Alfred P.\ Sloan Foundation (R.J.F), the David and Lucile Packard Foundation (R.J.F.\ and E.R.) and the Niels Bohr Professorship from the DNRF (E.R.).
A.M.B.\ acknowledges support from a UCMEXUS-CONACYT Doctoral Fellowship. W.H.L. is supported in part by UNAM-PAPIIT grant IG100317.

%
Support for this work was provided by NASA through Hubble Fellowship grant HST--HF--51373.001 awarded by the Space Telescope Science Institute, which is operated by the Association of Universities for Research in Astronomy, Inc., for NASA, under contract NAS5--26555.
This paper includes data gathered with the 6.5 meter Magellan Telescopes located at Las Campanas Observatory, Chile.
This research has made use of the NASA/IPAC Extragalactic Database (NED) which is operated by the Jet Propulsion Laboratory, California Institute of Technology, under contract with the National Aeronautics and Space Administration.
Based on observations made with the NASA/ESA Hubble Space Telescope, obtained from the Data Archive at the Space Telescope Science Institute, which is operated by the Association of Universities for Research in Astronomy, Inc., under NASA contract NAS 5--26555. These observations are associated with programs GO--14840.

\bibliography{gw.bib}
\end{document}